\documentclass{aastex631}

\begin{document}

\title{Bounding the Photon Mass with Ultrawide Bandwidth Pulsar Timing Data and Dedispersed Pulses of Fast Radio Bursts}

\author[0000-0002-9061-6022]{Yu-Bin Wang}
\affiliation{School of Physics and Electronic Engineering, Sichuan University of Science \& Engineering, Zigong 643000, China.\\}

\author[0000-0003-4686-5977]{Xia Zhou}
\affiliation{Xinjiang Astronomical Observatory, Chinese Academy of Sciences, Urumqi 830011, China.}
\affiliation{Key Laboratory of Radio Astronomy, Chinese Academy of Sciences, Nanjing 210008, China.}
\affiliation{Xinjiang Key Laboratory of Radio Astrophysics, 150 Science1-Street, Urumqi 830011, China.}

\author[0000-0002-2162-0378]{Abdusattar Kurban}
\affiliation{Xinjiang Astronomical Observatory, Chinese Academy of Sciences, Urumqi 830011, China.}
\affiliation{Key Laboratory of Radio Astronomy, Chinese Academy of Sciences, Nanjing 210008, China.}
\affiliation{Xinjiang Key Laboratory of Radio Astrophysics, 150 Science1-Street, Urumqi 830011, China.}

\author[0000-0003-4157-7714]{Fa-Yin Wang}
\affiliation{School of Astronomy and Space Science, Nanjing University, Nanjing 210093, China.}
\affiliation{Key Laboratory of Modern Astronomy and Astrophysics (Nanjing University), Ministry of Education, Nanjing 210093, China.}

\correspondingauthor{Xia Zhou}
\email{zhouxia@xao.ac.cn}



\begin{abstract}
Exploring the concept of a massive photon has been an important area in astronomy and physics. If photons have mass, their propagation in nonvacuum space would be affected by both the nonzero mass $m_{\gamma}$ and the presence of a plasma medium. This would lead to a delay time proportional to $m_{\gamma}^2\nu^{-4}$, which deviates from the classical dispersion relation (proportional to $\nu^{-2}$). For the first time, we have derived the dispersion relation of a photon with a nonzero mass propagating in plasma. To reduce the impact of variations in the dispersion measure (DM), we employed the high-precision timing data to constrain the upper bound of the photon mass. Specifically, the DM/time of arrival (TOA) uncertainties derived from ultrawide bandwidth (UWB) observations conducted by the Parkes Pulsar Timing Array (PPTA) are used. The dedispersed pulses from fast radio bursts (FRBs) with minimal scattering effects are also used to constrain the upper bound of photon mass. The stringent limit on the photon mass is determined by uncertainties of the TOA of pulsars, with an optimum value of $9.52\times 10^{-46} \, \rm kg \,\,(5.34 \times 10^{-10}\, \rm eV/c^2$). In the future, it is essential to investigate the photon mass, as pulsar timing data are collected by PTA and UWB receivers, or FRBs with wideband spectra are detected by UWB receivers.
\end{abstract}



\section{Introduction}\label{Sect.1}

In accordance with Maxwell's electromagnetism and Einstein's special relativity, it is assumed that all electromagnetic waves travel at a constant speed, $c$, in a vacuum \citep{Einstein1905}. This constant velocity implies that the photons do not have mass \citep{Weinberg1995}. If photons were to be found to have a nonzero mass, this would contradict the prediction and could lead to new physics. Therefore, testing the mass of photons ($m_{\gamma}$) using various methods across different scales, from particle physics to cosmology, is an interesting topic in modern physics. Examples of such tests include the null tests of Coulomb's inverse square law \citep{Williams1971}, measuring the Schumann resonance in a cavity \citep{Kroll1971}, examining Jupiter's magnetic field \citep{Davis1975}, surveying magnetohydrodynamic phenomena of the solar wind \citep{Ryutov1997,Ryutov2007,Retino2016}, observing the spin down of a white dwarf pulsar \citep{Yang2017}, studying the spin of supermassive black holes \citep{Pani2012}, and measuring the gravitational deflection of massive photons \citep{Lowenthal1973}.

The photon mass can be evaluated through the dispersion method, which involves analyzing the signals of cosmological transients at different frequencies to detect any potential frequency-dependent (chromatic) velocity caused by photons with a nonzero mass. This velocity variation would lead to frequency-dependent arrival times during observations (see review by \cite{Wei2021} and references therein). The dispersion method is used to identify frequency-dependent time delays from cosmological distances at different frequencies or energy bands, thus providing bounds on the photon mass. Radio pulse emissions of millisecond durations from sources such as radio pulsars and fast radio bursts (FRBs) are used to establish the upper limit of the photon mass, as shown in Table \ref{photon_mass_bound}.

\begin{table}
  \centering
  \caption{The stringent constraints on the photon mass have been determined through observations of both Galactic and Extragalactic sources.}\label{photon_mass_bound}
  \begin{center}
  \begin{tabular}{lllc}
  \hline\hline\noalign{\smallskip}
  Reference                                  &    Source (s)    & Bandwidth      
 &  $m_{\gamma}$ (kg)   \\
  \hline\noalign{\smallskip}
    Galactic source      &        &                      &  \\
   \cite{Lovell1964}   &      flare stars      &  0.54 $\mu$m$-$1.2 m & $1.6\times10^{-45}$ \\
   \cite{Warner1969}   &  Crab Nebula pulsar   &  0.35$-$0.55 $\mu$m & $5.2\times10^{-44}$ \\
    \cite{Chang2023}   &  Crab pulsar       &  $9.25$ GHz  &  $5.7\times10^{-46}$  \\
   This work   &   22 millisecond pulsars in Milky Way  &  0.7$\sim$4.0 GHz        & $9.52 \times 10^{-46}$   \\
    Extragalactic source      &         &                      &  \\
   \cite{Schaefer1999} &  GRB 980703           &  $5.0\times10^9-1.2\times10^{20}$ Hz& $4.2\times10^{-47}$ \\
   \cite{Zhang2016}    &  GRB 050416A          &  $8.46$ GHz$-$15 keV& $1.1\times10^{-47}$ \\
    \cite{Wei2017}     &  Extragalactic radio pulsar (PSR J0451-67)     &  $\sim 1.4$ GHz& $2.0\times10^{-48}$ \\
    \cite{Wei2017}     &  Extragalactic radio pulsar (PSR J0045-7042)      &  $\sim 1.4$ GHz& $2.3\times10^{-48}$ \\
    \cite{Wei2018}     &  22 radio pulsars in the LMC &  $\sim 1.4$ GHz  &  $1.5\times10^{-48}$  \\
    \cite{Wei2018}     &  5 radio pulsars in the SMC  &  $\sim 1.4$ GHz  &  $1.6\times10^{-48}$  \\
    \cite{Wu2016}      &  FRB 150418                  &  $1.2-1.5$ GHz  &  $5.2\times10^{-50}$  \\
    \cite{Bonetti2016} &  FRB 150418                  &  $1.2-1.5$ GHz  &  $3.2\times10^{-50}$  \\
    \cite{Bonetti2017} &  FRB 121102                  &  $1.1-1.7$ GHz  &  $3.9\times10^{-50}$  \\
    \cite{Shao2017}    &  21 FRBs (20 of them without redshift test) &  $\sim$ GHz  &  $8.7\times10^{-51}$  \\
    \cite{Xing2019}    &  FRB 121102 subbursts &  $1.34-1.37$ GHz  &  $5.1\times10^{-51}$  \\
    \cite{Wei2020}     &  9 localized FRBs &  $\sim$ GHz  &  $7.1\times10^{-51}$  \\
    \cite{Wang2021}    &  129 FRBs &  $\sim$ GHz  &  $3.1\times10^{-51}$  \\
    \cite{Lin2023}     &  17 localized FRBs &  $\sim$ GHz  &  $7.1\times10^{-51}$  \\
    \cite{Wang2023}     &  23 localized FRBs &  $\sim$ GHz  &  $3.8\times10^{-51}$  \\
    \cite{Chang2023}   &  FRB 180916.J0158+65   &  $110-188$ MHz  &  $6.0\times10^{-47}$  \\
   This work   &   3 localized nonrepeaters (NE2001)  &  1.10$\sim$1.44 GHz & $(4.52-4.72) \times 10^{-44}$   \\
   This work   &   3 localized nonrepeaters (YMW16)  &  1.10$\sim$1.44 GHz        & $(4.70-4.93) \times 10^{-44}$   \\
  \noalign{\smallskip}\hline\hline
  \end{tabular}
  \end{center}
\end{table}

Generally, the pulse arrival times of radio sources follow the classical dispersion relation ($\Delta t_{\rm d}\propto \nu^{-2}$) for the propagation of radio signals through a cold plasma. A constraint on the photon mass $m_{\gamma}$ is usually derived from the delay time with the accepted first-order Taylor expansion term ($\Delta t_{m_{\gamma},1}\propto m_{\gamma}^2\nu^{-2}$; \citealt{Wu2016,Chang2023}). However, the second-order delay time term ($\Delta t_{m_{\gamma},2}\propto m_{\gamma}^4\nu^{-4}$) has been used to limit the photon mass \citep{Chang2023}. This method produces upper limits that are 4 orders of magnitude higher (or worse) than those obtained from the first-order term \citep{Chang2023}. As the classical dispersion relation is calculated based on the speed of light for zero photon mass, it is hard to differentiate between the two dispersions, which leads to anomalous results when considering the second-order term. In this study, our aim is to address this problem by considering the new first-order
effects of delay time ($\Delta t_{m_\gamma}^{\prime}\propto m_{\gamma}^2\nu^{-4}$) that are caused by the combination of a nonzero photon mass and the plasma medium.

The delay time caused by the nonzero photon mass is an intrinsic characteristic of light that should be preserved in the dedispersed pulse of a radio source across high and low frequencies. Additionally, the geometric contribution to the delay time ($\Delta t_{\rm g}$), which is induced by the frequency-dependent propagation path, is proportional to $\nu^{-4}$ \citep{Er2020,Wang2022} and changes over time \citep{Cordes2017,Wang2022}, can significantly affect the dedispersed pulse, thus masking the delay time caused by the non-zero photon mass and leading to an increased upper limit of the photon mass. To reduce the geometric effects, two types of radio sources are observed to set an upper limit on the photon mass. Radio sources, such as Galactic pulsars and extragalactic FRBs, can be used to detect the delay time caused by the nonzero photon mass. The delay time of each dedispersed pulse between high and low frequencies in Galactic pulsars may be preserved in the time-independent white noise, which can be identified through the use of high-precision pulsar timing data and reflected in the time of arrival (TOA) uncertainties of pulsars. This delay time is mainly affected by the properties of photons with a nonzero mass and is significantly increased when radio pulses experience minimal scattering effects and the propagation path of the cosmological distance. Therefore, FRBs are also a suitable source for testing the photon mass.

\cite{Agazie2023} reported the results of pulsar timing using data from various radio telescopes and receivers. However, the bandwidths at the center frequency were found to be nonuniform, which could lead to larger TOA and dispersion measure (DM) uncertainties. To address this issue, \citet{Curylo2023} proposed wideband timing measurements to simultaneously estimate TOAs and DMs, along with their uncertainties and initial noise analysis. This method can provide comprehensive information on dispersive delay and different sources of timing noise, as well as system-independent white noise. Furthermore, \citet{Macquart2020} highlighted the importance of studying localized FRBs to differentiate the DM contributions from the intergalactic medium and the host galaxy.

This paper will explore the upper limit of the photon mass by using data from millisecond pulsars in the Milky Way, which were observed with the Ultrawide Bandwidth receiver by the Parkes Pulsar Timing Array (PPTA; \citealt{Curylo2023}. Additionally, the observed results of dedispersed radio pulses from certain localized FRBs, as provided by \cite{Day2020}, will be incorporated. The paper is structured as follows. Section \ref{sec2} will present the theoretical framework that takes into account the nonzero photon mass and plasma effects, from which the limits of $m_{\gamma}$ can be determined. Section \ref{sec3} will analyze the pulsar timing data of millisecond pulsars and dedispersed radio pulses from certain localized FRBs to set an upper limit for the photon mass. Finally, Section \ref{sec4} will discuss and conclude the findings.

\section{Theoretical Framework}\label{sec2}

\subsection{Dispersion relation combined with the nonzero photon mass and plasma effects}

Previous research has studied the delay time caused by a nonzero photon mass as a separate factor \citep{Wu2016,Shao2017,Wei2021}. The classical dispersion relation is derived from the alteration in the speed of light due to plasma, assuming that the photon has zeromass \citep{Wu2016,Shao2017}. Therefore, the total delay time is usually calculated by adding the delay times caused by the two independent dispersions. This implies that it is difficult to differentiate between the two effects and to obtain the maximum limit of the photon mass through direct observations \citep{Shao2017}. We looked into how radio signals with a nonzero photon mass are affected by the plasma in space. This will be discussed in the following.

The speed of light depends on the frequency, expressed as $\upsilon_{\rm g}^{\gamma} = c\sqrt{1 -A\nu^{-2}}$, where $A = m_{\gamma}^2 c^4/h^2$ and $h$ is the Planck constant. When comparing the group velocities of radio signals at different frequencies with the speed of light in a vacuum, it is essential to take into account the combined effects of plasma and the nonzero photon mass, rather than treating them as distinct entities.
Therefore, the group velocity can be expressed as
\begin{eqnarray}\label{photon_V}
\upsilon_{\rm g}^{\rm new} = \upsilon_{\rm g}^{\gamma} n_{\rm ref},
\end{eqnarray}
where $n_{\rm ref}$ is the index of refraction and given as \citep{Stix1992}
\begin{eqnarray}\label{refractive_index}
n_{\rm ref} = \left(1 - \frac{\nu_{\rm p}^2}{\nu^2} \right)^{1/2},
\end{eqnarray}
where $\nu_{\rm p} = (n_{\rm e}e^2/\pi m_{\rm e})^{1/2}$ is the plasma frequency, $n_{\rm e}$ represents the electron number density of the plasma medium along the line of sight, $e$ and $m_{\rm e}$ denote the charge and mass of an electron, respectively. Therefore, the observed delay time should be $\Delta t_{\rm obs} = \Delta t_{\rm d} + \Delta t_{m_\gamma}^{\prime}$ instead of $\Delta t_{\rm obs} = \Delta t_{\rm d} + \Delta t_{{m_\gamma}}$ ($\Delta t_{{m_\gamma}} \propto a_1 m_{\gamma}^2\nu^{-2} + a_2 m_{\gamma}^4\nu^{-4}$). 

From the classical dispersion relation, the dispersive delay is derived by the comparison between the time of monochrome light in the propagation of plasma medium and vacuum \citep{Lorimer2012}, so it is reasonable for the zero mass photon that the dispersive delay difference is caused due to different group velocities of light between the low and high frequencies in the plasma medium. However, if the photon had mass, the group velocity of the photon in vacuum would be $\upsilon_{\rm g}^{\gamma}$ rather than a constant $c$. Thus, assuming that two massive photons with different frequencies ($\nu_{\rm low}, \nu_{\rm high}$; $\nu_{\rm low} < \nu_{\rm high}$) are emitted at the same time from a radio source in the Milky Way and travel a distance of $d$ to reach the observer, the dispersive delay between the two photons (or two radio signals) can be expressed as
\begin{eqnarray}\label{delay_time_n}
\nonumber  \Delta t_{\rm d}^{\prime} &=& \int_0^{d(z_{\rm d})}\left(\frac{1}{\upsilon_{\rm g,low}^{\rm new}} -  \frac{1}{\upsilon_{\rm g, low}^{\gamma}} \right) {\rm d}l - \int_0^{d(z_{\rm d})} \left(  \frac{1}{\upsilon_{\rm g,high}^{\rm new}}- \frac{1}{\upsilon_{\rm g, high}^{\gamma}}\right){\rm d}l \\
                &\simeq& \frac{e^2}{2 \pi m_{\rm e} c} \left(\frac{1}{\nu_{\rm low}^{2}} - \frac{1}{\nu_{\rm high}^{2}}\right) {\rm DM} + \frac{Ae^2}{4 \pi m_{\rm e} c} \left(\frac{1}{\nu_{\rm low}^{4}} - \frac{1}{\nu_{\rm high}^{4}}\right) {\rm DM} + O(\nu)\\
\nonumber   &=& \Delta t_{\rm d}+ \Delta t _{m_\gamma}^{\prime} + O(\nu),
\end{eqnarray}
where $O(\nu)$ is the high-order delay time, which is not considered in this case; $\Delta t _{m_\gamma}^{\prime}$ is the delay time arises from the combined effects of plasma and nonzero photon mass and is written as
\begin{eqnarray}\label{delay_time_p2}
\Delta t _{m_\gamma}^{\prime} = \frac{Ae^2}{4 \pi m_{\rm e} c} \left(\frac{1}{\nu_{\rm low}^{4}} - \frac{1}{\nu_{\rm high}^{4}}\right){\rm DM}.
\end{eqnarray}

For the radio pulses from the extragalactic source at redshift $z_{\rm d}$, the dispersive delay, taking into account the results given by \cite{Deng2014}, can be expressed as
\begin{eqnarray}\label{delay_time_n1}
\nonumber  \Delta t_{\rm d}^{\prime} &\simeq& \frac{e^2}{2 \pi m_{\rm e} c} \left(\frac{1}{\nu_{\rm low}^{2}} - \frac{1}{\nu_{\rm low}^{4}}\right) {\rm DM} + \frac{Ae^2}{4 \pi m_{\rm e} c} \left(\frac{1}{\nu_{\rm high}^{2}} - \frac{1}{\nu_{\rm high}^{4}}\right)\\
& & \times\left[ {\rm DM_{\rm MW}} + {\rm DM_{\rm halo}} + \int_0^{d (z_{\rm d})}\frac{n_{\rm e}(z^{\prime}){\rm d}l}{(1+z^{\prime})^3} + \frac{\rm DM_{\rm host}}{(1+z_{\rm d})^3}\right] + O(\nu)\\
\nonumber   &=& \Delta t_{\rm d}+ \Delta t _{m_\gamma}^{\prime} + O(\nu),
\end{eqnarray}
where $z^{\prime}$ is the redshift and
\begin{eqnarray}\label{delay_time_p3}
\Delta t _{m_\gamma}^{\prime} = \frac{Ae^2}{4 \pi m_{\rm e}c} \left(\frac{1}{\nu_{\rm low}^{4}} - \frac{1}{\nu_{\rm high}^{4}}\right) {\rm DM_{\gamma}},
\end{eqnarray}
where $\rm DM_{\rm MW}$, $\rm DM_{\rm halo}$, and $\rm DM_{\rm host}$ are DMs contributed by the Milky Way, the Galactic halo, and the host galaxy, respectively. $n_{\rm e}(z^{\prime})$ represents the electron density from the Galaxy to the source (or in the intergalactic space) and $\rm DM_{\gamma}$ is an equivalent DM
\begin{eqnarray}\label{DM}
{\rm DM_{\gamma} } = {\rm DM_{\rm MW}} + {\rm DM_{\rm halo}} + \int_0^{d(z_{\rm d})} \frac{n_{\rm e}(z^{\prime}){\rm d}l}{(1+z^{\prime})^3} + \frac{\rm DM_{\rm host}}{(1+z_{\rm d})^3}.
\end{eqnarray}

The delay time caused by a nonzero photon mass is proportional to $m_{\gamma}^2\nu^{-4}$, which is different from the results of earlier studies \citep[$\propto m_{\gamma}^2\nu^{-2}$]{Wu2016,Shao2017}. Compared to Equation (\ref{delay_time_p2}), Equation (\ref{delay_time_p3}) implies that the photon mass constrained by extragalactic transient radio sources may be more uncertain, due to factors such as ${\rm DM_{\rm halo}}$, $n_{\rm e}(z^{\prime})$, and $\rm DM_{\rm host}$ \citep{Prochaska2019,Macquart2020,Cook2023}. The Milky Way's ${\rm DM_{\rm MW}}$ can be estimated using the NE2001 \citep{Cordes2002} or YMW16 model \citep{Yao2017}, though some uncertainty is introduced due to the uncertain electron density models within the Milky Way. Therefore, the photon mass would be more accurately determined by radio sources in the Galaxy than by radio pulses from extragalactic distances unless the uncertainties, such as $\rm DM_{\rm halo}$, $n_{\rm e}(z^{\prime})$, and $\rm DM_{\rm host}$, are taken into account.

\subsection{Residual Time delay caused by nonzero photon mass}

The total time delay of two photons with different frequencies can be attributed to several components, such as the geometric delay $\Delta t_{\rm g} \propto \nu^{-4}$ caused by the geometry and radiation processes of the radio source itself \citep{Wang2019,Tuntsov2021}, the effects of plasma lensing in the propagation path \citep{Cordes2016,Er2020,Wang2022,WangY2023}, and an unknown associated delay time $\Delta t_{\rm sys}$ with the observing systems of pulsars \citep{Shannon2014} or FRBs \citep{Gajjar2018}. This can be expressed as 
\begin{equation}
    \Delta t_{\rm obs} = \Delta t_{\rm g} + \Delta t_{\rm d} + \Delta t _{m_\gamma}^{\prime} + \Delta t_{\rm sys}
\end{equation}

For a pulsar located in the Milky Way, the uncertainties in TOA ($\sigma_{\rm TOA}$) due to the white noise from the observing system and analysis may include the delay time caused by nonzero photon mass and $\Delta t_{\rm sys}$. The geometric delay is possibly variable with the time due to the variations of DM and hidden in the red noise of the pulsar timing data. Therefore, the rest delay time is calculated as 
\begin{equation}
    \Delta t_{\rm rest}^{\rm Gal} = \Delta t_{\rm obs} - \Delta t_{\rm g} - \Delta t_{\rm d} = \Delta t _{m_\gamma}^{\prime} + \Delta t_{\rm sys}
\end{equation}

Assuming $\Delta t_{\rm sys} > 0$, then $\Delta t _{m_\gamma}^{\prime} = \Delta t_{\rm rest}^{\rm Gal} - \Delta t_{\rm sys} \leq \Delta t_{\rm rest}^{\rm Gal}$, which sets an upper boundary on the photon mass
\begin{eqnarray}\label{pulsar}
m_{\gamma} \leq hc^{-2} \left[\frac{4\pi m_{\rm e} c}{e^2} \frac{\Delta t_{\rm rest}^{\rm Gal}}{\left(\nu_{\rm low}^{-4} - \nu_{\rm high}^{-4}\right){\rm DM}} \right]^{1/2},
\end{eqnarray}
which can be simplified as
\begin{eqnarray}\label{pulsar1}
m_{\gamma} \leq (1.62\times 10^{-43}\, {\rm kg}) \left [\frac{\Delta t_{\rm rest,\mu s}^{\rm Gal}}{\left( \nu_{\rm low,GHz}^{-4} -\nu_{\rm high,GHz}^{-4}\right) {\rm DM}_{\rm pc \, cm^{-3}}} \right ]^{1/2},
\end{eqnarray}
where $\Delta t_{\rm rest,\mu s}^{\rm Gal}$ is normalized with $1 \rm \,\mu s$, $\nu_{\rm low,GHz}$ and $\nu_{\rm high,GHz}$ are normalized with $\rm{1 GHz}$, ${\rm DM}_{\rm pc \, cm^{-3}}$ is the normalized DM with ${1 \, \rm pc \, cm^{-3}}$.

For a radio transient source from a cosmic distance with redshift $z_{\rm d}$, the residual delay time of the signal between high and low frequencies after the single pulse has been dedispersed is represented by both $\Delta t_{\rm g}$ and $\Delta t_{m_{\gamma}}^{\prime}$. The rest delay time should be
\begin{eqnarray}\label{delay_time_rest1}
\nonumber\Delta t_{\rm rest}^{\rm extr} &=& \Delta t_{\rm obs} - \Delta t_{\rm d} \\
         &=& \Delta t _{m_\gamma}^{\prime} + \Delta t_{\rm g} + \Delta t_{\rm sys}.
\end{eqnarray}

If $\Delta t_{\rm g} + \Delta t_{\rm sys} > 0$ and the source has a relatively clean line-of-sight path, then $\Delta t _{m_\gamma}^{\prime} = \Delta t_{\rm rest}^{\rm extr} - \Delta t_{\rm g} - \Delta t_{\rm sys} \leq \Delta t_{\rm rest}^{\rm extr}$.
We can simplify calculations by approximating the effective DM from Equation (\ref{DM}) as follows:
\begin{eqnarray}\label{DM1}
{\rm DM_{\gamma}} \gtrsim {\rm DM_{\rm MW}} + {\rm DM_{\rm halo}} + \frac{{\rm DM_{\rm IGM} + DM_{\rm host}/(1 + z_{\rm d})}}{(1 + z_{\rm d})^2},
\end{eqnarray}
where ${\rm DM_{\rm IGM}}$ is the DM of intergalactic medium. Thus, the photon mass is constrained by considering the observed results of the extragalactic source
\begin{eqnarray}\label{FRB}
m_{\gamma} \leq hc^{-2} \left\{\frac{4\pi m_{\rm e} c}{e^2} \frac{\Delta t_{\rm rest}^{\rm extr}}{\left(\nu_{\rm low}^{-4} - \nu_{\rm high}^{-4}\right)} \left[ {\rm DM_{\rm MW}} + {\rm DM_{\rm halo}} + \frac{{\rm DM_{\rm IGM} + DM_{\rm host}/(1 + z_{\rm d})}}{(1 + z_{\rm d})^2}\right]^{-1}\right\}^{1/2},
\end{eqnarray}
which can be further reduced to
\begin{eqnarray}\label{FRB1}
m_{\gamma} \leq (1.62\times 10^{-43}\, {\rm kg}) \left\{\frac{\Delta t_{\rm rest, \mu s}^{\rm extr}}{\left( \nu_{\rm low,GHz}^{-4} -\nu_{\rm high,GHz}^{-4}\right )} \left[{\rm DM_{\rm MW}} + {\rm DM_{\rm halo}} + \frac{{\rm DM_{\rm IGM} + DM_{\rm host}/(1 + z_{\rm d})}}{(1 + z_{\rm d})^2}\right]^{-1}_{\rm pc \, cm^{-3}}\right\}^{1/2},
\end{eqnarray}
where $\Delta t_{\rm rest, \mu s}^{\rm extr}$ is normalized with $1\, \rm \mu s$, and ${\rm DM_{\rm MW}}, {\rm DM_{\rm halo}}, {\rm DM_{\rm IGM}}, {\rm DM_{\rm host}}$ are normalized  with ${\rm 1 pc \, cm^{-3}}$. Since the numerator of the third term on the right side of Equation (\ref{DM1}) can be easily estimated by $\rm DM_{\rm tot} - DM_{\rm MW} - DM_{\rm halo}$, where $\rm DM_{\rm tot}$ is the total DM of a FRB. This method by comparing to the previous works is not affected by the uncertainties of
$\rm DM_{\rm host}$ and $\rm DM_{\rm IGM}$ \citep{Zhang2020,Zhang2021}.

\section{Photon mass limits}\label{sec3}

Generally, the discrepancies between the timing model and the observations of a pulsar's TOA can be seen as two types of timing noise: red and white noise \citep{Lorimer2012}. The former may include time-dependent effects, such as pulsar intrinsic spin noise, variations in DM, gravitational waves, and so on. On the other hand, the latter does not depend on the observed time \citep{Curylo2023}. The delay time caused by nonzero photon mass may be present in the white noise of pulsar timing data, which is reflected in the TOA uncertainties. 

FRBs are bright, extragalactic, transient radio pulses with a duration of a few milliseconds. The TOA of bursts at different frequencies may show a slight deviation from the classical dispersion relation \citep{Thornton2013}. This
discrepancy could potentially be caused by the nonzero photon mass. Consequently, FRBs can be employed to limit the photon mass.

\subsection{Timing data of 22 millisecond pulsars}\label{subsec3.1}

Recently, \cite{Curylo2023} reported the timing results of 35 millisecond pulsars that were observed using a UWB low-frequency receiver installed on the 64 m Parkes radio telescope. This receiver provided instantaneous frequency coverage in the range of 0.704-4.032$\rm \, GHz$. However, some pulsars that had either no or only one eigenprofile had a low flux measured above 2.5$\rm \, GHz$, and their bands above $3-3.5\, \rm GHz$ were excluded \citep{Curylo2023}. Therefore, we chose data of 22 millisecond pulsars with more than two high signal-to-noise ratio (S/N) eigenprofiles as reported by \cite{Curylo2023}. The data in Table \ref{WBP} include the DM and the median uncertainties of the TOA  ($\sigma_{\rm TOA}$) which can be thought of as the results of white noise. It is clear that the PPTA observations are highly precise, with an uncertainty of TOA as low as $10\, \rm ns$ for PSR J1939$+$2134. With the timing results of all millisecond pulsars in Table \ref{WBP} and $\Delta t_{\rm rest}^{\rm Gal} = \sigma_{\rm TOA}$, the upper limits of the photon mass, as given by Equation (\ref{pulsar1}), are provided in the last column of Table \ref{WBP}. The limits of the photon mass range from $2.48\times 10^{-44} \, \rm kg$ (or $1.39\times 10^{-8}\, \rm eV/c^2$) to $9.52\times 10^{-46} \, \rm kg$ (or $5.34 \times 10^{-10}\, \rm eV/c^2$). Since the nonzero photon mass is an inherent property, there exists a strict upper limit for its mass 
\begin{eqnarray}\label{photon_mass}
m_{\gamma} \leq 9.52\times 10^{-46} \, \rm kg \simeq 5.34 \times 10^{-10}\, \rm eV/c^2.
\end{eqnarray}

The observing strategy and precision of the radio telescope, the frequency-dependent delay caused by the nonzero photon mass and the cold plasma medium, and the fluctuations of the interstellar medium can all have an effect on the DM perturbations of radio pulses \citep{Cordes2010, Curylo2023}. These factors can lead to an uncertainty in the DM of a radio pulse. We assume that the DM uncertainty is mainly due to the nonzero photon mass, and thus Equation (\ref{pulsar1}) can be used to calculate the upper limit of the photon mass. \citealt{Curylo2023} reported the median uncertainties of the DM ($\sigma_{\rm DM}$) of millisecond pulsars at the epoch length of the timing observations, which are presented in Table \ref{WBP}. It was found that the limits on the photon mass are slightly higher when the median uncertainties of the DM of pulsars are taken into account, compared to those calculated using the median uncertainties of the TOA. The best result from the example of PSR J1939$+$2134 yielded an upper limit of $m_{\gamma} \leq 1.92\times 10^{-45}\rm \, kg \simeq 1.08\times 10^{-9} \, eV/{c^2}$. It is clear that the uncertainties of the DM have a moderate effect on the limits of the photon mass.

\begin{table*}[!ht]
  \centering
  \caption{The data obtained from PPTA \citep{Curylo2023} and the calculated photon mass in this work (last two column). The PPTA, using a UWB receiver ($0.704 -4.032\rm \, GHz$), obtained timing results (${\rm DM, median \ \sigma_{\rm TOA}, median \ \sigma_{\rm DM}}$) for 22 millisecond pulsars with more than two high-S/N edge profiles \citep{Curylo2023}.}\label{WBP}
  \begin{center}
  \begin{tabular}{lcccccccc}
  \hline\hline\noalign{\smallskip}
  No.    & Pulsars         &  DM                  & Med. $\sigma_{\rm TOA}$ & Med. $\sigma_{\rm DM}$                 &$m_{\gamma}^{\rm TOA}$ &$m_{\gamma}^{\rm DM}$    \\
        &                 &  ($\rm pc\, cm^{-3}$)&    ($\mu$s)            & ($\times 10^{-4}\, \rm pc\, cm^{-3}$) &    (kg)               &    (kg)                 \\
  \hline\noalign{\smallskip}
  01    &PSR J0613$-$0200 & 38.78                &      0.158             &      0.807                            & $5.12\times10^{-45}$  & $1.04\times10^{-44}$   \\
  02    &PSR J0711$-$6830 & 18.41                &      0.550             &      3.211                            & $1.39\times10^{-44}$  & $3.02\times10^{-44}$  \\
  03    &PSR J0900$-$3144 & 75.61                &      0.612             &      3.421                            & $7.22\times10^{-45}$  & $1.54\times10^{-44}$  \\
  04    &PSR J1017$-$7156 & 94.22                &      0.120             &      0.684                            & $2.86\times10^{-45}$  & $6.16\times10^{-45}$  \\
  05    &PSR J1024$-$0719 & 6.48                 &      0.619             &      4.105                            & $2.48\times10^{-44}$  & $5.76\times10^{-44}$  \\
  06    &PSR J1045$-$4509 & 58.15                &      0.781             &      4.173                            & $9.30\times10^{-45}$  & $1.94\times10^{-44}$  \\
  07    &PSR J1125$-$6014 & 52.93                &      0.119             &      0.744                            & $3.81\times10^{-45}$  & $8.57\times10^{-45}$  \\
  08    &PSR J1545$-$4550 & 68.39                &      0.212             &      2.144                            & $4.47\times10^{-45}$  & $1.28\times10^{-44}$  \\
  09    &PSR J1600$-$3053 & 52.33                &      0.113             &      0.832                            & $3.73\times10^{-45}$  & $9.12\times10^{-45}$  \\
  10    &PSR J1603$-$7202 & 38.04                &      0.361             &      2.129                            & $7.82\times10^{-45}$  & $1.71\times10^{-44}$  \\
  11    &PSR J1643$-$1224 & 62.41                &      0.286             &      1.459                            & $5.43\times10^{-45}$  & $1.11\times10^{-44}$  \\
  12    &PSR J1713$+$0747 & 15.92                &      0.048             &      0.384                            & $4.41\times10^{-45}$  & $1.12\times10^{-44}$  \\
  13    &PSR J1730$-$2304 & 9.62                 &      0.287             &      1.677                            & $1.39\times10^{-44}$  & $3.02\times10^{-44}$  \\
  14    &PSR J1744$-$1134 & 3.14                 &      0.080             &      0.524                            & $1.28\times10^{-44}$  & $2.95\times10^{-44}$  \\
  15    &PSR J1857$+$0943 & 13.30                &      0.257             &      2.189                            & $1.12\times10^{-44}$  & $2.93\times10^{-44}$  \\
  16    &PSR J1939$+$2134 & 71.01                &      0.010             &      0.050                            & $9.52\times10^{-46}$  & $1.92\times10^{-45}$  \\
  17    &PSR J2051$-$0827 & 20.73                &      0.680             &      5.657                            & $1.45\times10^{-45}$  & $3.78\times10^{-43}$  \\
  18    &PSR J2145$-$0750 & 9.00                 &      0.186             &      1.179                            & $1.15\times10^{-44}$  & $2.62\times10^{-44}$  \\
  19    &PSR J2241$-$5236 & 11.41                &      0.070             &      0.465                            & $6.29\times10^{-45}$  & $1.46\times10^{-44}$  \\
  20    &PSR J0125$-$2327 & 9.60                 &      0.129             &      0.974                            & $9.30\times10^{-45}$  & $2.30\times10^{-44}$  \\
  21    &PSR J1022$+$1001 & 10.25                &      0.254             &      1.517                            & $1.26\times10^{-44}$  & $2.78\times10^{-44}$  \\
  22    &PSR J0437$-$4715 & 2.64                 &      0.005             &      0.043                            & $3.49\times10^{-45}$  & $9.23\times10^{-45}$  \\
  \noalign{\smallskip}\hline\hline
  \end{tabular}
  \end{center}
  \begin{enumerate}
      \item[] \textbf{Note}. The upper limit of the photon mass, $m_{\gamma}^{\rm TOA}$ and $m_{\gamma}^{\rm DM}$, was calculated from the median $\sigma_{\rm TOA}$ and $\sigma_{\rm DM}$, respectively.
    \end{enumerate}
\end{table*}

\subsection{Dedispersed Pulses of three localized FRBs}

FRBs are divided into two categories: one-off events (nonrepeaters) and repeating bursts (repeaters). More than 20 FRBs have been identified, along with their locations and host galaxies \citep{Macquart2020,Niu2022,Caleb2023,Driessen2023}. Repeaters have much higher DM in their host galaxy than in intergalactic space \citep{Niu2022,Wu2024}. The degree of linear polarization for repeaters is significantly dependent on frequency \citep{Feng2022,Wang2022a}. These studies suggest that bursts from a repeater at different frequencies may come from different propagation paths, which can be attributed to the presence of an inhomogeneous plasma medium \citep{Er2020,Feng2022,Wang2022,WangY2023,Yang2023}. This phenomenon leads to a considerable geometric delay in the dedispersion of bursts between low and high frequencies \citep{Er2020,Wang2022}. Furthermore, there is no obvious evidence that nonrepeaters have spectral depolarizations \citep{Uttarkar2023}. Therefore, we select nonrepeaters with a scattering timescale lower than $1\,\rm ms$ at $1\,\rm GHz$ to reduce the geometric effects in the propagation path \citep{Er2020}, and neglect the geometric delay.

FRBs can be dedispersed in two ways: by maximizing the average S/N across the band to reach its peak value, or by optimizing the alignment of the frequency-time structure \citep{Gajjar2018}. The difference in the DM between the two methods ($\Delta \rm DM$) is due to the delay time of $t_{m_{\gamma}}^{\prime} \propto \nu^{-4}$ caused by the existence of a nonzero photon mass. So the rest delay time is taken as $\Delta t^{\rm extr}_{\rm rest} \approx 4.15\, {\rm ms}\times \Delta \rm DM(\nu^{-2}_{\rm low, GHz} - \nu^{-2}_{\rm high,GHz})$. Table \ref{T_photon_mass} presents the data of three non-repeaters detected by the Australian Square Kilometer Array Pathfinder (ASKAP; $1.1045- 1.4405\,\rm GHz$), including the names of FRBs, the DMs from the de-dispersed method of the optimized alignment of frequency-time structure, $\Delta \rm DM$, and $\rm DM_{\rm MW}$ from the electron density model of NE2001 \citep{Cordes2002} or YMW16 \citep{Yao2017}. The spectrum of three FRBs appears in all bands of ASKAP. Furthermore, the observation of FRB in M81 shows $\rm DM_{\rm halo} \lesssim 32-42 \, \rm pc \, cm^{-3}$ \citep{Kirsten2022}. CHIME/FRB detected a nearby FRB with DM of no more than 250 pc cm$^{-3}$, leading to the $\rm DM_{\rm halo}$ range of 52-111 pc cm$^{-3}$ \citep{Cook2023}.  X-ray observations give $\rm DM_{\rm halo} \lesssim 10 \, \rm pc \, cm^{-3}$ \citep{Keating2020}. Thus, we take the possible $\rm DM_{\rm halo}$ in the range of 0-111 pc cm$^{-3}$. By combining the parameters in Table \ref{T_photon_mass}, $\rm DM_{\rm halo}$, and Equation (\ref{FRB1}), the upper limits of the photon mass can be calculated
\begin{eqnarray}\label{photon_mass1}
m_{\gamma} \leq (4.52 - 4.92)\times 10^{-44} \, \rm kg \simeq (2.54 - 2.76)\times 10^{-8}\, \rm eV/c^2
\end{eqnarray}
for the NE2001 model, and 
\begin{eqnarray}\label{photon_mass2}
m_{\gamma} \leq (4.70 - 5.16)\times 10^{-44} \, \rm kg \simeq (2.64 - 2.90)\times 10^{-8}\, \rm eV/c^2
\end{eqnarray}
for the YMW16 model. The photon mass can only be estimated with the reliable electron density models from the Milky Way and the potential contributions of different types of DM from the Galactic halo. This makes it difficult to accurately determine the photon mass using extragalactic sources.

\noindent\hfill
\begin{table*}
  \centering
  \caption{The data of three FRBs observed by ASKAP \citep{Day2020} and the calculated photon mass in this work (last two column). The parameters $\rm{z_{\rm d}, DM, \Delta DM}$ of three FRBs were obtained from observations made by ASKAP \citep{Day2020}. }\label{T_photon_mass}
  \begin{center}
  \begin{tabular}{lccccccccc}
  \hline\hline\noalign{\smallskip}
  Source        & $z_{\rm d}$ &   DM                  & $\Delta \rm DM$      & $\rm DM_{\rm MW}^N$  & $\rm DM_{\rm MW}^Y$   & $m_{\gamma}^{\rm N}$ &  $m_{\gamma}^{\rm Y}$ \\
                &             &  ($\rm pc\, cm^{-3}$) & ($\rm pc\, cm^{-3}$) & ($\rm pc\, cm^{-3}$) & ($\rm pc\, cm^{-3}$)  &  ($\times 10^{-44}\,$kg)  
                       &  ($\times 10^{-44}\,$kg)                 \\
  \hline\noalign{\smallskip}
   FRB 180924   &  0.3212     & 362.16                &  0.04                & 40.5                 & 27.65                 & $10.08-12.19$ & $11.19-12.34$ \\
   FRB 190102   &  0.2912     & 364.545               &  0.007               & 57.3                 & 3.48                  & $4.52-4.92$  & $4.70-5.16$   \\
   FRB 190611   &  0.3778     & 332.63                &  0.03                & 57.83                & 3.07                  & $9.91-11.12$  & $10.46-11.91$  \\
  \noalign{\smallskip}\hline\hline
  \end{tabular}
  \end{center}
  \begin{enumerate}
      \item[] \textbf{Note}. The upper limits of the photon mass, denoted by the superscripts `N' and `Y', were calculated by the NE2001 \citep{Cordes2002} and YMW16 \citep{Yao2017} models, respectively.
  \end{enumerate}
\end{table*}

\section{Conclusions and discussion}\label{sec4}

In this paper, we present a revised dispersion relation for photons with nonzero mass as they travel through a plasma medium. This updated relation, denoted as $t_{m_{\gamma}}^{\prime}\propto m_{\gamma}^2\nu^{-4}$, is different from the classical dispersion relation ($t_{\rm d}\propto \nu^{-2}$) and the previously established delay time proportional to $\propto m_{\gamma}^2 \nu^{-2}$. The intrinsic delay time of light, $t_{m_{\gamma}}^{\prime}$, is independent of the observed time and can be observed as DM/TOA uncertainties in pulsar timing data or as dedispersed radio pulses across a range of frequencies. In this study, we used high-precision pulsar timing data from the UWB receiver of PPTA and the dedispersed pulse of the FRB with minimal scattering effects to determine the upper limit of photon mass. The results of this work are listed below.
\begin{itemize}
    \item From Pulsar timing. The median uncertainties of the TOA from 22 millisecond pulsars were used to calculate the upper limit of the photon mass, which is $m_{\gamma} \leq 9.52\times 10^{-46} \, \rm kg $ ($m_{\gamma} \leq 5.34 \times 10^{-10}\, \rm eV/c^2$).
    \item From FRB. The upper limit of the photon mass was determined by considering the time delay between high and low frequencies for the dedispersed radio pulses of three localized FRBs. The calculated upper limit is $m_{\gamma} \leq (4.52-5.16)\times 10^{-44} \, \rm kg$ ($m_{\gamma} \leq (2.54-2.90) \times 10^{-8}\, \rm eV/c^2$). 
\end{itemize}

It should be noted that the pulsar timing data collected in the frequency range of 0.704$-$4.032$\rm \, GHz$ provide a more stringent constraint on the value of $m_{\gamma}$ than the delay time difference of the FRBs observed between 1.1045 and 1.4405$\, \rm GHz$. This increased precision is due to the fact that the dispersion relation caused by the presence of a photon mass is a higher-order term that depends on frequency. Thus, even after applying the classical dispersion relation to remove dispersion effects, the delay time associated with the photon mass can still be preserved in a radio pulse. This delay time is especially prominent at frequencies below $1 \,\rm{GHz}$. Consequently, it is plausible to suggest that the upper limit of the photon mass can be more accurately determined by analyzing data obtained from observations conducted at frequencies below $1 \,\rm{GHz}$.

We have given an upper bound for the photon mass in our investigation through a newly updated formula. This is the first time that the interaction between a nonzero photon mass and the plasma medium has been taken into account and calculated as the photon propagates through the plasma medium. Our estimated limit exceeds the limits set in prior studies ($m_{\gamma}\leq 10^{-47}\, \rm kg$ or $10^{-51}\, \rm kg$), which used the second-order dispersion relation or the first-order dispersion relation, by approximately 2 or 5 orders of magnitude, respectively. It is worth noting that these earlier limits were determined under the assumption of a photon with zero mass traveling through the plasma medium. Therefore, it may be necessary to revisit these limits once more by incorporating the factors we have developed in this work.

The upper limit of the photon mass given in our work is not much lower than the one given by \citealt{Chang2023}. This could be due to recent studies that provide only the median TOA uncertainties from long-term pulsar timing data, without considering the minimum white noise level of the pulsar \citep{Agazie2023,Curylo2023}. Additionally, the use of UWB receivers has been shown to increase the S/N ratio in recorded pulsar profiles, thus reducing the uncertainty in measuring pulse TOA \citep{Liu2014}. Comparison of noise models from narrowband and wideband datasets can help distinguish between different sources of noise \citep{Alam2021, Agazie2023, Curylo2023}. Taking advantage of highly sensitive radio facilities such as the Five-hundred-meter Aperture Spherical Radio Telescope \citep{Jiang2019} and the Qitai Radio Telescope \citep{Wang2023SCPMA}, along with the implementation of UWB receivers, there is potential for further improvement in determining the upper limit of photon mass in the near future.

Although we take these nonrepeaters with short scattering timescales in order to reduce the geometric effects acting on the TOA of bursts, the chosen FRBs still provide an upper limit for the photon mass that is about ten times higher than the values obtained from ultrawideband pulsar timing data. This suggests that some physical mechanisms around the source or in the propagation pathway may be retained in the dedispersed pulses of bursts yet. According to the observed results, the spectra of many nonrepeaters have a broader range than the full CHIME/FRB band ($400-800 \rm \, MHz$), implying the improved TOA accuracy of nonrepeater detections by using multiple facilities simultaneously or UWB receivers \citep{Pleunis2021}. In addition, the effecitive DM ($\rm DM_{\gamma}$) included the term of photon mass is an approximate value here. In future studies, the upper limit of the photon mass may be revised by the dedispersed pulses of nonrepeaters if the contributions of these physical mechanisms on the TOA of bursts is eliminated and the estimated method of $\rm DM_{\gamma}$ is improved.


\section{Acknowledgments}
\begin{acknowledgments}{}
We thank the anonymous referee for value suggestions that can allow us to improve our manuscript significantly. This work is supported in part by the National Key R\&D Program of China (No. 2022YFC2205202), the Natural Science Foundation of Xinjiang Uygur Autonomous Region (No. 2023D01E20, 2022D01A363), the National Natural Science Foundation of China (No. 12033001, 12288102, 12273028, 12273009), the Sichuan Science and Technology Program (No. 24QYCX0261, 2023YFG0024, 4GJHZ0046), the Sichuan University of Science \& Engineering Program (No. 2022RC07), the Opening Foundation of Xinjiang Key Laboratory (No. 2021D04016), the Major Science and Technology Program of Xinjiang Uygur Autonomous Region (No. 2022A03013-1, 2022A03013-4), and the Urumqi Nanshan Astronomy and Deep Space Exploration Observation and Research Station of Xinjiang (XJYWZ2303). XZ would like to thank Qing-min Li for her fruitful suggestions. 
\end{acknowledgments}


\end{document}